Alexander J. Tybl

# An Overview of Spatial Econometrics

**Abstract**

This paper offers an expository overview of the field of spatial econometrics. It first justifies the necessity of special statistical procedures for the analysis of spatial data and then proceeds to describe the fundamentals of these procedures. In particular, this paper covers three crucial techniques for building models with spatial data. First, we discuss how to create a spatial weights matrix based on the distances between each data point in a dataset. Next, we describe the conventional methods to formally detect spatial autocorrelation – both global and local. Finally, we outline the chief components of a spatial autoregressive model, noting the circumstances under which it would be appropriate to incorporate each component into a model. This paper seeks to offer a concise introduction to spatial econometrics that will be accessible to interested individuals with a background in statistics or econometrics.

**Introduction**

Spatial econometrics, as its name suggests, is the subfield of econometrics dedicated to analyzing spatial data. The necessity for specialized techniques to analyze spatial data arises from the fact that, in general, spatial data violate the assumption that data must be independent. This lack of independence is formally known as spatial autocorrelation. In many ways, spatial econometrics is closely related to time series analysis, since it aims to address spatial autocorrelation similarly to how time series analysis aims to address temporal autocorrelation. However, there are also many important differences between spatial econometrics and times series analysis. This paper highlights several similarities and differences between these two branches of statistical modeling, whenever applicable.

Spatial econometrics deals primarily with three types of datasets. The first of these is cross-sectional data. A cross-sectional dataset consists of many observations, all recorded at the same time or during the same time period. The other two types are both examples of spatio-temporal data. Of these, the first is panel data, also known as longitudinal data. Panel data consists of various observations recorded over time of the same individuals. The third type, cross-sectional data pooled over time, is similar to panel data, except a dataset of this type does not necessary include observations of the same individuals at each time period.

The approach taken in this paper is inspired by that of Dubé and Legros' *Spatial Econometrics Using Microdata*. First, we discuss various ways to construct spatial weights matrices based on the distances between data points. Next, we summarize the conventional methods to detect spatial autocorrelation, both global and local. Finally, we explain how to incorporate both of these procedures into spatial autoregressive models.

**Spatial Weights Matrices**

What happens at one location affects all other locations, but, the closer the other location is to the first, the more it will be affected. Thus, in order to predict the value of a variable at a given location, we must not only know about the surrounding locations but also have a procedure for weighting the influence of circumstances at the surrounding locations.

To this end, the first step is to formally construct a method to represent distance. In spatial econometrics, we assume the locations of data exist on a Cartesian plane. However, in contrast to standard geometry, there are two prevalent yet notably different measures of distance in spatial econometrics. The first is the familiar Euclidean distance: Given two locations $i = (X_i, Y_i), j=(X_j, Y_j)$, the Euclidean distance between them is given by $d_{ij} = \sqrt{(Y_i - Y_j)^2 + (X_i - X_j)^2}$. The other measure of distance, known as the Manhattan distance, measures the length of a lattice path between two points: Given two locations $i = (X_i, Y_i), j=(X_j, Y_j)$, the Manhattan distance between them is given by $d_{ij}^\star = |Y_i - Y_j| + |X_i - X_j|$. The chief appeal of the Manhattan distance is that emulates the distance one would need to travel in a city grid system. However, we are interested in an entire dataset of locations, rather than only two. Thus, we construct a distance matrix

$$\mathbf{D} = \begin{pmatrix} 0 & d_{12} & \cdots & d_{1j} & \cdots & d_{1N} \\ d_{21} & 0 & \cdots & d_{2j} & \cdots & d_{2N} \\ \vdots & \vdots & \vdots & \vdots & \vdots & \vdots \\ d_{i1} & d_{i2} & \cdots & d_{ij} & \cdots & d_{iN} \\ \vdots & \vdots & \vdots & \vdots & \vdots & \vdots \\ d_{N1} & d_{N2} & \cdots & d_{Nj} & \cdots & 0 \end{pmatrix}$$ (if we are interested in Euclidean distance) or

$$\mathbf{D}^\star = \begin{pmatrix} 0 & d_{12}^\star & \cdots & d_{1j}^\star & \cdots & d_{1N}^\star \\ d_{21}^\star & 0 & \cdots & d_{2N}^\star & \cdots & d_{2N}^\star \\ \vdots & \vdots & \vdots & \vdots & \vdots & \vdots \\ d_{i1}^\star & d_{i2}^\star & \cdots & d_{ij}^\star & \cdots & d_{iN}^\star \\ \vdots & \vdots & \vdots & \vdots & \vdots & \vdots \\ d_{N1}^\star & d_{N2}^\star & \cdots & d_{Nj}^\star & \cdots & 0 \end{pmatrix}$$ (if we are interested in Manhattan distance).

Note that the diagonal entries of the distance matrix are zero, since the distance between any location and itself is always zero. Also, note that the distance matrix is symmetric, since the distance between locations $i$ and $j$ is always equal to the distance between locations $j$ and $i$.

Recall that, the smaller the distance between two locations is, the greater the locations' influence on one another will be. However, in the distance matrix, the smaller the distance between two locations is, the smaller the corresponding matrix entry will be (by definition). Thus, in order to incorporate the spatial autocorrelation into a model, it is necessary to come up with a transformation for the distance matrix so that, the smaller the distance between two locations is, the greater the corresponding transformed matrix entry will be. The transformed matrix is known as a spatial weights matrix and is denoted

$$\mathbf{W} = \begin{pmatrix} 0 & w_{12} & \cdots & w_{1j} & \cdots & w_{1N} \\ w_{21} & 0 & \cdots & w_{2j} & \cdots & w_{2N} \\ \vdots & \vdots & \vdots & \vdots & \vdots & \vdots \\ w_{i1} & w_{i2} & \cdots & w_{ij} & \cdots & w_{iN} \\ \vdots & \vdots & \vdots & \vdots & \vdots & \vdots \\ w_{N1} & w_{N2} & \cdots & w_{Nj} & \cdots & 0 \end{pmatrix}.$$

There is no standard or optimal method to construct the spatial weights matrix, but there are four particularly prevalent transformations, which we will review briefly.

One transformation, based on connectivity relations, produces a binary matrix with entries equal to 1 for locations with distance less than or equal to some distance threshold $\bar{d}$ and 0 otherwise. That is, $w_{ij} = \begin{cases} 1 \text{ if } d_{ij} \leq \bar{d} \; \forall i,j = 1, \cdots, N; i \neq j \\ 0 \text{ otherwise} \end{cases}$. Another transformation, based on inverse distance, is given by $w_{ij} = \begin{cases} d_{ij}^{-\gamma} \; \forall i,j = 1, \cdots, N \\ 0 \; \forall i = j \end{cases}$, where $\gamma$ is a parameter that can be specified by the individual creating the matrix. For example, $\gamma = 1$ yields the inverse distance, $\gamma = 2$ yields the inverse squared distance, etc. A third, inverse exponential-based, transformation is given by $w_{ij} = e^{-d_{ij}} \; \forall i \neq j; i,j = 1, \cdots, N$. A fourth method to construct the spatial weights matrix, known as Gaussian transformation, is given by $w_{ij} = \begin{cases} [1 - (\frac{d_{ij}}{\bar{d}})^2]^2 & \forall d_{ij} \leq \bar{d}; i \neq j; i,j = 1,\cdots, N \\ 0 & \forall d_{ij} > \bar{d} \\ 0 & \forall i = j \end{cases}$. Again, recall that there is no standard procedure for constructing a spatial weights matrix. In fact, it is even possible to combine different procedures. For example, one could combine the transformation based connectivity relations with that based on the inverse distance to produce the following transformation: $w_{ij} = \begin{cases} d_{ij}^{-\gamma} & \text{if } d_{ij} \leq \bar{d} \ \forall i,j = 1,\cdots,N \\ 0 & d_{ij} > \bar{d} \\ 0 & \forall i = j \end{cases}$.

Several econometricians have set forth guidelines regarding which transformations to use for different types of modeling problems. Chen (3) lays out the following guidelines:

- If the area of the geographical region of interest is large, then use the inverse distance-based transformation.
- If the area of the geographical region of interest is small, then use the inverse exponential-based transformation.
- If the suspected influence of other locations occurs primarily on a local scale, then use the connectivity-based transformation.

Regarding the distance threshold $\bar{d}$ in the connectivity-based transformation, Griffith (79) argues that it is better to set $\bar{d}$ equal to a value less than the true distance threshold than to set $\bar{d}$ equal to a value greater than the true distance threshold.

Once the spatial weights matrix has been constructed, it is convention to standardize the matrix so that all rows sum to one. Formally, we denote the standardized

$$\text{row matrix } \mathbf{W}^\star = \begin{pmatrix} 0 & w^\star_{12} & \cdots & w^\star_{1j} & \cdots & w^\star_{1N} \\ w^\star_{21} & 0 & \cdots & w^\star_{2N} & \cdots & w^\star_{2N} \\ \vdots & \vdots & \vdots & \vdots & \vdots & \vdots \\ w^\star_{i1} & w^\star_{i2} & \cdots & w^\star_{ij} & \cdots & w^\star_{iN} \\ \vdots & \vdots & \vdots & \vdots & \vdots & \vdots \\ w^\star_{N1} & w^\star_{N2} & \cdots & w^\star_{Nj} & \cdots & 0 \end{pmatrix}, \text{ where } w^\star_{ij} = \frac{w_{ij}}{\sum_{j=1}^{N} w_{ij}}.$$

**Spatial Autocorrelation**

While the (standardized) spatial weights matrix allows us to adjust for the effects of spatial autocorrelation, we must also have a procedure for verifying the presence and measuring the intensity of spatial autocorrelation. There are two types of spatial autocorrelation: global and local. Global spatial autocorrelation refers to the presence of spatial trends that cover the scale of the entire geographical region of interest, while local spatial autocorrelation refers to the presence of spatial trends that occur in pockets throughout the region of interest.

The most commonly used measurement of global spatial autocorrelation is Moran's $I$ statistic. Because it measures the strength of the linear relationship between the value of some variable $y$ at every location $i$ and the same variable at all other locations $j$ in the dataset, Moran's $I$ can be considered analogous to the correlation coefficient. By definition, Moran's $I$ is given by $I = \frac{N}{\sum_{i=1}^{N}\sum_{j=1}^{N} w_{ij}} \frac{\sum_{i=1}^{N}\sum_{j=1}^{N} w_{ij}(y_i - \bar{y})(y_j - \bar{y})}{\sum_{i=1}^{N}(y_i - \bar{y})^2}$.

In order to use Moran's $I$ for significance testing, $y$ must follow a normal distribution. If this is the case, then Moran's $I$ follows a normal distribution with mean $E[I] = -\frac{1}{N-1}$ and variance $Var[I] = \frac{N^2 \frac{\sum_{i=1}^{N}\sum_{j=1}^{N}(w_{ij}+w_{ji})^2}{2} - N\sum_{i=1}^{N}(w_{i\cdot}+w_{\cdot i})^2 + 3(\sum_{i=1}^{N}\sum_{j=1}^{N}w_{ij})^2}{\left(\sum_{i=1}^{N}\sum_{j=1}^{N}w_{ij}\right)^2(N^2-1)} - E[I]^2$.

A significance test for global spatial autocorrelation has as its null hypothesis $H_0: I = 0$ (absence of spatial autocorrelation) and can have alternative hypothesis $H_a: I \neq 0$, $H_a: I > 0$, or $H_a: I < 0$ (presence of spatial autocorrelation, positive spatial autocorrelation, or negative spatial autocorrelation, respectively). Once the appropriate hypotheses have been stated, the next step is to calculate $t = \frac{I - E[I]}{\sqrt{Var[I]}}$. After this step, the significance test proceeds as any other $t$-test. Ideally, we would fail to reject the null hypothesis. The presence of spatial autocorrelation can cause the estimated linear regression coefficients, as well as these coefficients' estimated variances, to become biased. We will address how to resolve these issues in the following section on spatial autoregression models.

For local spatial autocorrelation, there is an analogue to Moran's $I$ known as the Local Moran index $I_i$. By definition, $I_i = (y_i - \bar{y})\sum_{j=1}^{N}w_{ij}(y_j - \bar{y})$ for $i \neq j$. Like its global counterpart, the Local Moran index can only be used for significance testing if $y$ is normally distributed. If this condition holds, then $I_i$ has mean $E[I_i] = \frac{-\sum_{j=1}^{N}w_{ij}}{N-1}$ and variance $Var[I_i] = \frac{\sum_{=1,j\neq i}^{N} w_{ij}^2 (N - \frac{\sum_{i=1}^{N} y_i^{\star 4}}{N}/(\frac{\sum_{i=1}^{N} y_i^{\star 2}}{N})^2)}{N-1} + \frac{2\sum_{k\neq i}\sum_{h\neq i} w_{ik}w_{ih}(2\frac{\sum_{i=1}^{N} y_i^{\star 4}}{N}/\left(\frac{\sum_{i=1}^{N} y_i^{\star 2}}{N}\right)^2 - N)}{(N-1)(N-2)} - \frac{(\sum_{j=1}^{N} w_{ij})^2}{(N-1)^2}$.

When testing for local spatial autocorrelation, Anselin (96) proposes adjusting the significance level according to sample size so that the significance threshold becomes $\frac{\alpha}{N}$, rather than $\alpha$. Regardless, any significance test for local spatial autocorrelation must be interpreted cautiously. Contrary to the global Moran's $I$, which follows a normal distribution, the distribution of the local Moran index is unknown, although Boots and Tiefelsdorf (333) proved that the local Moran index does not follow a normal distribution.

**Spatial Autoregressive Models**

Similarly to time series analysis, spatial autoregression models can include a lagged specification on independent variables, on dependent variables, or on the error term. Here, we will review the circumstances under which it would be advantageous to include each of these and how to incorporate them into a model.

The first circumstance we will consider deals with the presence of externalities. Externalities occur when an individual location benefits or suffers from a characteristic(s) of nearby locations, but the individual location does not have the ability to influence the characteristic(s). An example of externalities would be for a location to benefit somehow from the nearby presence of a beautiful beach. Externalities are often present in endogenous growth models and in new economic geography. If one suspects externalities to be present in some dataset of interest, the proper way to address this in a model would be to introduce a lag on the independent variables. Formally, such a model would take on the form $y_i = \beta_0 + \beta_1 x_{i1} + \cdots + \beta_K x_{iK} + \gamma_1 x_{j1} + \cdots + \gamma_K x_{jK} + \varepsilon_i$, or, in matrix form, $\mathbf{y} = \mathbf{X}\beta + \mathbf{W}\mathbf{X}\gamma + \varepsilon$. Here, the $\beta_k$ coefficients measure the impact of a unit change in $x_{ik}$ (for location $i$) on $y_i$, all else constant, whereas the $\gamma_k$ coefficients measure the impact of a unit change in $x_{jk}$ (for location $j$) on $y_i$, all else constant. In order for this model to be valid, its residuals must have mean zero, exhibit homoscedasticity, and be independent of each other (absence of spatial autocorrelation). The first two of these conditions can be verified exactly as how one would verify them in any other multiple linear regression model. The independence of residuals can be checked using Moran's $I$. Assuming these conditions are satisfied, it holds that $\frac{\partial \mathbf{y}}{\partial \mathbf{X}} = \mathbf{I}\beta + \mathbf{W}\gamma$, where $\mathbf{I}$ is the $N \times N$ identity matrix.

The second circumstance we will consider is the presence of spillover effects. We say that spillover effects are present if the value of the dependent variable at a given location is influenced by the values of the dependent variable at nearby locations. One example of a situation in which spillover effects are likely present is when the price of a house is influenced by the prices of nearby houses. The presence of spillover effects in spatial econometrics can be considered analogous to the presence of dynamic effects in time series analysis. Thus, if one suspects spillover effects, then it would be advantageous to introduce a lag on the dependent variable of the model. Formally, such a model would take on the form $y_i = \rho y_j + \alpha + \beta_1 x_{i1} + \cdots + \beta_K x_{iK} + \varepsilon_i$, or, in matrix form, $\mathbf{y} = \mathbf{W}\mathbf{y}\rho + \mathbf{X}\beta + \varepsilon$. Here, the $\rho$ coefficient measures the average influence of the values of the dependent variable at other locations on the value of the dependent variable at location $i$. The coefficients in this model can be estimated using either the generalized method of moments or the maximum likelihood method. As for the previous model, order for this model to be valid, its residuals must have mean zero, exhibit homoscedasticity, and be independent of each other (absence of spatial autocorrelation). If these conditions are satisfied, then it holds that $\frac{\partial \mathbf{y}}{\partial \mathbf{X}} = (\mathbf{I} - \mathbf{W}\rho)^{-1}\mathbf{I}\beta$.

The third circumstance that we will consider is spatial heterogeneity. We say spatial heterogeneity occurs when the third condition (absence of spatial autocorrelation) for each of the two previous models fails. Often, spatial heterogeneity is a result of omitted variable bias for some spatially structured lurking variable. Sometimes, it is possible to solve this issue immediately by including the lurking variable in the model. However, it is often impossible to include the lurking variable in the model, especially when the lurking variable is difficult to quantify. In this situation, it is possible to address the issue by including a lag on the error term in the model. Formally, such a model would take on the form $y_i = \alpha + \beta_1 x_{i1} + \cdots + \beta_K x_{iK} + v_i$, where $v_i = v_j \lambda + \varepsilon_i$ is an error term that includes the influence $\lambda$ of the lurking variable $v_j$ on $y_i$. The parameter $\lambda$ is estimated to ensure that $\varepsilon_i$ is independent of other error terms. Although the derivation of the following equation is beyond the scope of this report, it is important to note that the resulting model will take on the form $y_i = \alpha + \beta_1 x_{i1} + \cdots + \beta_K x_{iK} + (1-\lambda)^{-1}\varepsilon_i$, or, in matrix form, $\mathbf{y} = \mathbf{X}\beta + (\mathbf{I} - \lambda \mathbf{W})^{-1}\varepsilon$. As in the previous model, the coefficients in this model can be estimated using either the generalized method of moments or the maximum likelihood method. Because the independence of error terms is automatically satisfied in this approach, the only two conditions that remain to be satisfied in order for this model to be valid are that the residuals have mean zero and exhibit homoscedasticity. Assuming these conditions are satisfied, it holds that $\frac{\partial \mathbf{y}}{\partial \mathbf{X}} = \mathbf{I}\beta$, just as in regular multiple linear regression.

While the three modeling techniques outlined in this section are presented separately, it is important to note that they can be applied in conjunction with each other. For instance, if one is working with a dataset in which they suspect the presence of both externalities and spillover effects, they should include lagged specifications on both the dependent variable and the independent variables in the model.  Additionally, although it would be beyond the scope of this report to go into detail regarding significance tests for the utility of the model, it should still be mentioned that there are three prominent statistical tests to this purpose in the context of spatial autoregressive models.  These include the likelihood ratio test, the Wald test, and the Lagrange multiplier test.

**Conclusion**

Given the substantial importance of spatial data in economics, as well as countless other disciplines, it is crucial that there exist sound statistical techniques to analyze such data.  It is somewhat surprising, therefore, that this branch of statistical modeling has not attracted considerable attention until recent decades.  Nevertheless, in this time, numerous econometricians have thoroughly developed the subtleties of spatial econometrics.  All of this scholarly development, however, is based on the same foundations that this paper attempts to broadly outline.  Among these are the ideas of using distance to construct spatial weights matrices, detecting and measuring global and local spatial autocorrelation, and building spatial autoregressive models.  In terms of autoregressive modeling techniques, spatial econometrics and time series analysis are exceedingly similar.  However, the more fundamental procedures of constructing spatial weights matrices and measuring and detecting spatial autocorrelation are, in contrast, fairly unique to spatial econometrics.